\documentclass[english,aps,prx,amsmath,twocolumn,amssymb,nofootinbib,superscriptaddress,showpacs,floatfix,longbibliography]{revtex4-2}

\usepackage[english]{babel}

\usepackage[utf8]{inputenc}
\usepackage{latexsym}
\usepackage{braket}	
\usepackage{amsmath,amssymb,amsfonts,textcomp,latexsym,pb-diagram,amsopn,bm}
\usepackage{graphicx,xcolor}
\graphicspath{{pictures/}}
\DeclareGraphicsExtensions{.pdf,.png,.jpg,.eps}
\usepackage[breaklinks,colorlinks,bookmarks=false,citecolor=blue,linkcolor=red,urlcolor=blue]{hyperref}
\usepackage{url}
\usepackage{hyperref}
\usepackage{tabularx}
\newcolumntype{L}{>{\raggedright\arraybackslash}X}
\usepackage{epstopdf}

\begin{document}

\title{Restrictions on the existence of weak values in quantum mechanics: \\ weak quantum evolution concept}

\author{Gleb A. Skorobagatko}

\affiliation{%
\mbox{%
Institute for Condensed Matter Physics of National Academy of Sciences of Ukraine,
  Svientsitskii Str.1,79011 Lviv, Ukraine%
  }}

\email{ <gleb.a.skor@gmail.com>}

\begin{abstract}
\begin{small}
It is shown, that the Aharonov-Albert-Vaidman concept of weak values appears to be a consequence of a more general quantum phenomenon of \textit{weak quantum evolution}. Here the concept of weak quantum evolution is introduced and discussed for the first time. In particular, it is shown on the level of quantum evolution that there exist restrictions on the applicability of weak quantum evolution- and, hence, weak values approach. These restrictions connect the size of a given quantum ensemble with the parameters of pre- and post-selected quantum states. It is shown, that the latter requirement can be fulfilled for the model system, where the concept of weak values was initially introduced by Aharonov, Albert and Vaidman. Moreover, the deep connection between weak quantum evolution and conventional probability of quantum transition between two non-orthogonal quantum states is established for the first time. It is found that weak quantum evolution of quantum system between its two non-orthogonal quantum states is inherently present in the measurement-determined definition of quantum transition probability between these two quantum states.  
\end{small}
\end{abstract}

\maketitle
\begin{normalsize}

\section{Introduction}

Among numerous peculiarities of quantum mechanical description of nature the simple concept of operator weak values as quantum system's characterization between any two of its quantum states is one of the most questionable. All the story has began in 1988, when Aharonov, Albert and Vaidman have shown in their seminal (AAV-) paper\cite{1} , that if one transforms system's initial quantum state $ \vert i\rangle $ by acting on it with certain  operator $ \hat{A} $ and then projects the result on certain other quantum state  $ \vert f\rangle $, which is non-orthogonal to  $ \vert i\rangle $ then the resulting projection  $ \langle f \vert\hat{A} \vert i\rangle $  divided by the non-zero overlap of these two states $\langle f \vert i\rangle $ will be equal to what they define as \textit{"weak value"} $ A_{w}= \frac{\langle f \vert \hat{A} \vert i\rangle}{\langle f \vert i\rangle } $ of operator $ \hat{A} $ with respect to two given quantum states $ \vert i\rangle $  and $ \vert f\rangle $. These two non-orthogonal quantum states can be referred to as \textit{initial}, or \textit{pre-selected} and \textit{final}, or \textit{post-selected} quantum states of quantum system being disturbed by acting on it with operator  $ \hat{A}  $  during the time interval between the moments of system pre- and post-selection\cite{1}.

Within such framework, the disturbance caused by operator $ \hat{A} $ while acting on given quantum system can always be interpreted as a sort of \textit{measurement} performed on given quantum system between its pre- and post-selection. In such context pre- and post-selection should be understood as procedures performed on the ensemble of identical quantum systems before and after such the ensemble would experience a "measurement" by the operator $ \hat{A} $.  Obviously, in  the situation where $ \vert i\rangle=\vert f\rangle $ weak value $ A_{w} $ reduces to the conventional average $ \langle i \vert\hat{A} \vert i\rangle $ of operator $ \hat{A} $ with respect to given (pre-selected) quantum state of the system and in the particular case when $ \vert i\rangle $ is an eigenstate of operator $ \hat{A} $ such average is by definition just the corresponding eigenvalue $ A_{i} $ from the spectrum of operator $ \hat{A} $ . 

At this point one immediately sees the main peculiarity of weak values as compared to conventional quantum averages: in most general case pre- and post-selected states being non-orthogonal to each other can still have very small overlap $ \vert \langle f \vert i\rangle \vert \ll 1 $ which means that weak value $ A_{w} $ of operator $ \hat{A} $ can be much larger than any of its eigenvalues. In this case one has \textit{anomalous weak value}. Evidently, the realization of such situation strongly depends on the particular choice of three independent characteristics of given quantum system: on the form of system disturbance (or measurement) i.e. on the form of operator $ \hat{A} $ as well as on two chosen (pre- and post-selected) quantum states of given quantum system. Therefore, the main achievement of Aharonov, Albert and Vaidman in their AAV-paper was  the successful choice of these free characteristics for realistic quantum system they had chosen. 
 
This way, during the recent thirty years, both in theory \cite{2}  and related experiments \cite{3,4,5}  the concept of weak values has got a sense of independent characteristic of those quantum systems experienced measurement of certain their characteristics (spin projections or photon polarizations) between the procedures of their pre- and post-selections(see e.g. Refs.[2,3,4]). Moreover, recently it has been understood and confirmed experimentally for finite ensemble of polarized photons \cite{5} that quantum weak values though being introduced for ensembles of quantum systems have more common features with operator eigenvalues rather than with conventional ensemble averages \cite{5}.  On the other hand, as one could notice already from a very general definition of weak value in the above, the \textit{time arrow} is inherently present in \textit{any} experimentally relevant definition of weak values\cite{1,2,3,4,5}. 

Therefore, such type of quantum mechanical averages as weak values, obviously, can serve as certain additional and quite exotic marker of system's evolution from its "initial" to "final" state, while disturbed by measurement operator $ \hat{A} $ in between. Taking into account this observation, Aharonov and Vaidman in their later works  on the subject (see e.g. Ref.[2] and references therein) had concluded that weak value of quantum operators can represent even more fundamental quantum object than a conventional concept of operator eigenvalues does, incorporating the latter only as important particular case\cite{2}. This has led them further to the concept of two-vector formalism being an interesting philosophical generalization of Dirac's "bra-" and "ket" vector of states quantum description\cite{2}. Another  approach developed independently by A.Matzkin in \cite{11} also concerns weak values as certain additional properties of quantum systems those being complementary to ordinary eigenvalues in certain types of weak measurements.  Thus, weak values have emerged as technically conventional but quite paradoxical by nature example of quantum mechanical description, having caused numerous experimental and theoretical proposals\cite{2,3,4,5}. Naturally, precisely because of that a lot of attempts to trivialize the entire weak value concept and underlying approach by refuting the strangeness of anomalous weak value phenomenon have been done. The most radical criticism was intended to refer  the entire effect of anomalous weak values just to exotic post-selection procedures on the classical ensembles which have nothing to do with quantum physics\cite{6}. Whereas another studies \cite{7,8}  tried to refer all the phenomena related to weak values to the artefacts of strong quantum fluctuations during the incompatible projective measurements performed on the non-commuting observables very close to each other in the time domain. However, despite or due to such polar opinions on the true meaning of weak values, that concept still remains a mysterious peculiarity in the body of conventional quantum mechanics people still trying to understand and agree on.

So far, within the above-mentioned context, the aim of this paper is to give a clear, natural and at the same time, qualitatively novel quantum statistical explanation to weak value phenomena, including anomalous weak values situation, within the standard ideology of quantum mechanics all members of the community could agree on. Especially, below I will show, in which way the statistical meaning of weak values is connected with the projective measurement-accompanied evolution of quantum system, what are the restrictions on the ensemble size under which the evolution of each quantum system of the ensemble will be characterized by weak value of observable and finally what is the quantum statistical role of anomalous weak values in definition of quantum transition probabilities.

\section{Weak values of evolution operator and quantum transition probability}

Below I will keep most general theoretical description taking into account the possibility of widest interpretation of related results, while the parallel with historically first model system by Aharonov, Albert and Vaidman \cite{1} will be also maintained. Therefore, let us begin with most common definition of weak value \cite{1}

\begin{equation} \label{1}
A_{w}= \frac{\langle f \vert \hat{A} \vert i\rangle}{\langle f \vert i\rangle },
\end{equation}

where quantities $\hat{A}$, $ \vert i\rangle $ and $ \vert f\rangle $ have been already defined in the above. Now let us use this expression in a bit inconvenient way, not as widely known definition of  weak value, but as a \textit{new definition of transition amplitude from pre- to post-selected quantum state via weak measurement resulting in weak value}

\begin{equation} \label{2}
\langle f \vert i\rangle = \frac{\langle f \vert \hat{A} \vert i\rangle}{A_{w}}.
\end{equation}

One can see that if weak value is defined by Eq.(1) than Eq.(2) contains no new information as compared to Eq.(1). However, below we will see that it actually does if one takes Aharonov-Albert-Vaidman (AAV-) theoretical model in its most general fashion. Beforehand, let us make quite straightforward trick, multiplying both sides of Eq.(2) on projection $ \langle i \vert f\rangle $ (which is always non-zero by the assumption). Taking into account standard quantum mechanical definition of the transition probability $ P_{i\rightarrow f}=\vert \langle f \vert i\rangle \vert^{2}= \langle i \vert f\rangle\langle f \vert i\rangle$ between two arbitrary quantum states $ \vert i\rangle $ and $ \vert f\rangle $ as well as common definition of the projection operator $ \hat{\Pi}_{f}=\vert f\rangle \langle f \vert$ on the state $ \vert f \rangle $ one obtains

\begin{equation} \label{3}
P_{i\rightarrow f} = \frac{\langle i \vert \hat{\Pi}_{f} \hat{A} \vert i\rangle}{\langle i \vert\hat{\Pi}_{f}\hat{A}\vert i\rangle_{w}},
\end{equation}

where we have introduced a new notation for weak value as follows $ \langle i \vert\hat{\Pi}_{f}\hat{A}\vert i\rangle_{w}= A_{w}$ the latter means that standard definition (1) of operator weak value encodes a special procedure of averaging in which projections of the sort $ \hat{\Pi}_{f}=\vert f\rangle \langle f \vert$ are involved, this procedure will be discussed in details in what follows.  Equation (3) already contains  something interesting enough since it defines common quantum transition probability in a new way, \textit{as the ratio between two different types of measurement performed on a given quantum system} (remember that operator $ \hat{A} $ as well as any projection operator $ \hat{\Pi}_{f}=\vert f\rangle \langle f \vert$ both can be associated with certain type of measurement on given quantum system). Namely, as it was initially introduced in Ref.[1], if $ \hat{A} $ represents certain continuous (but not necessary weak) quantum measurement between pre- and post-selection, while post-selection $ \hat{\Pi}_{f} $ itself is represented by another continuous quantum measurement, which follows the measurement performed by $ \hat{A} $, then the numerator of Eq.(3) is the average over initial (pre-selected) quantum state of two operators $ \hat{A} $ and $ \hat{\Pi}_{f} $ acting on the system consequently. It is important that the  numerator in the r.h.s. of Eq.(3) involves just one act of projective measurement described by $ \hat{\Pi}_{f} $, whereas the denominator in the r.h.s. of Eq.(3) represents something more complicated. In what follows we will see that in most general situation where weak values can appear the denominator of Eq.(3) contains \textit{all possible sequences} of special projective measurements each representing a definite modification of  $ \hat{\Pi}_{f} $. To see this in details, let us turn to the situation have been discussed in Ref.[1].

\medskip

As it was shown first in AAV-paper\cite{1}, in order to obtain weak value of arbitrary operator $ \hat{A} $ one should consider the quantum amplitude $ \langle f \vert \hat{U}_{A} \vert i\rangle $ associated with system's evolution governed by evolution operator $ \hat{U}_{A} $ between the states  $ \vert i \rangle $  and $ \vert f \rangle $. Evolution operator $ \hat{U}_{A} $ describes how the interaction associated with  the process of continuous measurement of observable $ A $ changes initial (or pre-selected) quantum state of the system. In Ref.[1] and further related papers the simplest form of $ \hat{U}_{A} $ is chosen $ \hat{U}_{A}=\exp(-i\varepsilon_{A} q \hat{A}) $, where $  \hat{A} $ is an operator related to physical observable $ A $ of measured system, while variable $ q $ - characterizes coordinate of the detector (such a detector variable is usually counted as one being normally distributed at the stage of system pre-selection: $ P_{det}\approx \exp(-(q-q_{0})^{2}/4\Delta^{2}) $, where $ q_{0} $ -is the center of detector coordinate distribution, $ \Delta $ is its spread, while for the positive parameter $ \varepsilon_{A} $ in general one has $ \varepsilon_{A}=\varepsilon_{A}(t) $, i.e. it is a function of time. Obviously, operator $ \hat{U}_{A} $ describes system's evolution due to its interaction with detector, where corresponded interaction Hamiltonian is $ \hat{H}_{A}=\frac{d\varepsilon_{A}(t)}{dt}q \hat{A} $. Naturally, in AAV-paper\cite{1} for $ \hat{H}_{A} $ one has $ \hat{A}=\sigma_{z} $ , $ q=z $ and $ \frac{d\varepsilon_{A}(t)}{dt}=-\mu\left(\partial B_{z}/\partial z \right)g[y(t,\tau_{w})] $, where $\partial B_{z}/\partial z $ is a gradient of external magnetic field in $ z $-direction, while function $ g[y(t,\tau_{w})] \propto [\theta(t+\tau_{w})-\theta(t-\tau_{w})]$ is a "square pulse"-function of time ($t \geq 0$, $ \tau_{w}>0 $) where the width of the pulse $ \tau_{w} $ is the electron's "time of flight" in $ y $-direction through the region of a non-zero gradient of $ B_{z} $. 

Now since the nature of operator $ \hat{A} $ in Eq.[3] is not specified, i.e. it can be arbitrary, one can straightforwardly re-write Eq.(3) with operator $ \hat{U}_{A} $ instead of $ \hat{A} $. This results in a remarkable relation

\begin{equation} \label{4}
P_{i\rightarrow f} = \frac{\langle i \vert \hat{\Pi}_{f} \hat{U}_{A} \vert i\rangle}{\langle i \vert\hat{U}_{A}(\hat{\Pi}_{f})\vert i\rangle_{w}}.
\end{equation}  

Now it is time to notice that the main approximate transformation of Ref.[1] which has led authors of that paper to a weak value  $ A_{w} $ of operator $ \hat{A} $ is the replacement of a following exact time-propagator

\begin{eqnarray}\label{5} 
\frac{\langle f(t) \vert \hat{U}_{A}(t)\vert i\rangle}{\langle f \vert i \rangle}=\frac{\langle f(t)\vert \mathcal{T}_{t}\exp(-i\int_{0}^{t}dt' \hat{H}_{A}(t')) \vert i \rangle}{\langle f \vert i \rangle} 
\end{eqnarray}

by its weak-valued approximation 

\begin{eqnarray} 
 \langle f(t) \vert\hat{U}_{A}(t)\vert i\rangle_{w} = \exp \left( -i\int_{0}^{t}dt' \frac{\langle f(t') \vert \hat{H}_{A}(t')\vert i \rangle}{\langle f \vert i \rangle}\right) \nonumber
\end{eqnarray}

\begin{equation}
\label{6}
\end{equation}

here $ \langle f(t) \vert\hat{U}_{A}(t)\vert i\rangle_{w}= \exp(-i S_{w}(t))$ with $ S_{w}(t)=\int_{0}^{t}dt'\frac{\langle f(t') \vert \hat{H}_{A}(t') \vert i \rangle}{\langle f \vert i \rangle}=\varepsilon_{A} (t) q_{z} A_{w} $ represents a "weak action" for interaction Hamiltonian $ \hat{H}_{A} $ of the system. Such weak action is proportional to weak value $ A_{w} $ of operator $ \hat{A} $. Since the total evolution operator $ \hat{U}(t) $ of two subsequent measurements can be written as product $ \hat{U}(t)=\hat{U}_{s}(t)\hat{U}_{A}(t) $ in Eqs.(5,6) one can assign a part of total system's evolution due to $ \hat{U}_{s}(t) $ to the time evolution of a "final" state bra-vector $\langle f \vert $ in a way $\langle f(t) \vert=\langle f \vert \tilde{U}_{s}(t)  $, where $ \tilde{U}_{s}(t)=\exp(-i S_{st}(t)\hat{B})  $ is evolution operator with "strong" action $ S_{st}(t)=\int_{\tau_{w}}^{t}dt' \hat{H}_{s}(t') $ related to strong measurement (post-selection) with  corresponded Hamiltonian (of strong measurement) $ \hat{H}_{s}(t)=(d\varepsilon_{st}(t)/dt )q_{x} \hat{B} $. Here  the operator $ \hat{B} $ of strong measurement normally does not commute with operator $ \hat{A} $ of preceding weak measurement and $ d\varepsilon_{st}(t)/dt \propto [\theta(t+\tau_{s})-\theta(t-\tau_{s})]$ is also a "square pulse"-function of time with $ \tau_{s} $ being a characteristic time interval of a strong measurement which is subsequent to weak measurement of interest providing that $\tau_{s}>\tau_{w} $. As well, symbol $ \mathcal{T}_{t} $ in the definition (5) of system's evolution operator $ \hat{U}_{A}(t) $ means time-ordering operator acting on the time-dependent, operator-valued exponential operator. Notice, also that the "pulse-functions" of time for the interactions responsible for consequent weak and strong measurements of the ensemble under condition $\tau_{s}>\tau_{w} $ allows for the relation $  \mathcal{T}_{t}\hat{U}_{s}(t)\hat{U}_{A}(t)=\hat{U}_{s}(t)\mathcal{T}_{t}\hat{U}_{A}(t) $ for the time-ordering, as one can see also from Eqs.(5,6).  In the case of the AAV-model \cite{1} one has $ \hat{B}=\sigma_{x} $ while  $ \hat{A}=\sigma_{z} $, that is why classical variable $ q $ of the detector has different indices ($ x $ and $ y $) in definition of $ \hat{H}_{s}(t) $ and $ \hat{H}_{A}(t) $, correspondingly (see e.g. Ref.[1] for further details\cite{9} ). (Here and everywhere below we have put $ \hbar=1 $ for simplicity.)

Obviously, Eq.(5) represents exact statement, while Eq.(6) is nothing but the result of a certain approximation. Practically, in the most simple cases, such as e.g. one from AAV-paper\cite{1}, the latter approximation can be made either assuming $ \varepsilon_{A} \ll 1 $ or by taking a large enough ensemble of $ N \gg 1 $ identical quantum systems and performing a weak measurement with "ensemble-averaged" interaction Hamiltonian $ \tilde{H}_{A}=\frac{\hat{H}_{A}}{N} $ on each subsystem in the ensemble (see  below). In the latter situation it is presumed that $ N $ is so large that commutator $ [\tilde{H}_{A}(t_{1}),\tilde{H}_{A}(t_{2})] $ tends to zero for arbitrary values of $ t_{1} $ and $ t_{2} $,. However, these assumptions provide the validity of Eq.(6) only for a single quantum object (e.g. for one electron) while for the ensemble of $ N \gg 1 $ identical weakly measured quantum objects (e.g. for the beam of electrons being initially prepared in the same quantum state $ \vert i \rangle $ each) the validity of approximation (6) requires more complicated analysis. Especially, as it will be shown in the next section, the approximation (6) for the ensemble of $ N \gg 1 $ identical quantum particles breaks down when $ N $ is large enough. It is important to mention here that related condition (4) from Ref.\cite{1} is not satisfactory for such the analysis because condition (4) from \cite{1}  refers to the "cut off" of all higher orders in (5) due to the initial gaussian distribution of just one particle's coordinate with small enough spread $ \Delta $ while for the ensemble of $ N \gg 1$ identically prepared quantum particles one has evident renormalization $ \Delta \rightarrow \Delta\sqrt{N} $ for the initial spread of weakly measured observable. Thus, inequality (4) from \cite{1} breaks down in such the case for $ N \gg 1$.  Anyway, the approximation (6) turns out to be crucial for the entire concept of operator weak values. \textit{Therefore, one may think of Eq.(6) as of independent definition of "weak quantum evolution", the latter can be regarded as a special sort of quantum evolution associated with the existence of weak values of a given weakly measured observable between system's pre- and post-selections.} In what follows I will reveal and explain the basic distinctive features of a novel "\textit{weak quantum evolution}" (or simply, \textit{weak evolution}) phenomenon.

\section{Weak quantum evolution approach: limits of applicability}

Now, the limits of the applicability for the approximation (6) of "weak quantum evolution" should be clarified.  In order to establish to what extent the approximation of Eq.(6) can be valid in most general cases one needs to expand exact T-exponent in the Eq.(5)

\begin{eqnarray}\label{6}
\frac{\langle f(t) \vert \hat{U}_{A}(t)\vert i\rangle}{\langle f \vert i \rangle}=  1 -i\int_{0}^{t}dt_{1} \frac{\langle f(t_{1})\vert \tilde{H}_{A}(t_{1}) \vert i \rangle}{\langle f \vert i \rangle} \nonumber \\
+ (-i)^{2} \int_{0}^{t}dt_{1} \int_{0}^{t_{1}}dt_{2} \frac{\langle f(t_{1}) \vert \tilde{H}_{A}(t_{1}) \tilde{H}_{A}(t_{2}) \vert i \rangle}{\langle f \vert i \rangle}  \nonumber \\
+ \ldots +(-i)^{n} \int_{0}^{t}dt_{1} \ldots \nonumber \\
\ldots \int_{0}^{t_{n-1}}dt_{n} \frac{\langle f(t_{1})\vert \tilde{H}_{A}(t_{1}) \ldots \tilde{H}_{A}(t_{n}) \vert i \rangle}{\langle f \vert i \rangle}+\ldots \nonumber \\
\end{eqnarray}  

Obviously, analogous temporal evolution of $ N \gg 1$ identical quantum systems is described by the operator $ \hat{U}_{N,A}(t)=\hat{U}_{A}(t)^{\otimes N} $ where each operator $ \hat{U}_{A}(t) $ is exponential with the "ensemble-weighted" Hamiltonian $ \tilde{H}_{A}(t)=\frac{\hat{H}_{A}(t)}{N} $ instead of $ \hat{H}_{A}(t) $. Notice, that the latter procedure of the the "ensemble averaging" of the first (weak) measurement on the system is introduced here as obvious equivalent of the spread renormalization $ \Delta \rightarrow \Delta\sqrt{N} $ in the initial gaussian distribution of the observable $ A $ eigenvalues for the ensemble of $ N $ identical quantum subsystems all being prepared (or, alternatively, pre-selected) in the same initial quantum state $ \vert i \rangle $. In Eq.(7) the "ensemble-averaged" Hamiltonian $ \tilde{H}_{A}(t) $ acts on the product state $ \vert i_{N}\rangle=\sqcap_{j=1}^{N}\otimes\vert i_{j}\rangle $ and then the result becomes projected on the other (post-selected) product state $ \vert f_{N}(t)\rangle=\sqcap_{j=1}^{N}\otimes\vert f_{j}(t)\rangle $. Then series (5) can be written for each among the $ N $ identical quantum subsystems of the ensemble involving $N$ pre- and post-selected quantum states of subsystems being involved. (Therefore, without the loss of generality we will skip index $ j $ in what follows). Notice the time argument in post-selected vector of state $ \vert f_{N}(t)\rangle $, it is due to the non-zero Hamiltonian $ \hat{H}_{s}(t) $ associated with "strong" measurement and needed for system post-selection. As the result, for the ensemble of $ N $ identical quantum systems one obtains

\begin{equation} \label{7}
\langle f_{N}(t) \vert \tilde{U}^{\otimes N}_{A}(t)\vert i_{N}\rangle=\left( \langle f(t) \vert \tilde{U}_{A}(t)\vert i\rangle \right)^{N} 
\end{equation}

where $\tilde{U}_{A}(t) $ is from Eq.(7) with Hamiltonian $ \tilde{H}_{A}(t)=\hat{H}_{A}(t)/N $. Due to non-orthogonality of states $ \vert i\rangle $ and $ \vert f \rangle $ one can always decompose $  \vert i \rangle$ as follows $ \vert i \rangle=\vert f \rangle \langle f \vert i \rangle + \vert \bar{f} \rangle \langle \bar{f}\vert i \rangle $, where vectors of states $ \vert f \rangle $ and $ \vert \bar{f} \rangle $ form complete orthonormal basis of eigenstates for Hamiltonian $ \hat{H}_{A}(t) $ with properties: $ \langle \bar{f} \vert f \rangle =\langle f \vert \bar{f} \rangle=0$ and $ \langle f \vert f \rangle=\langle \bar{f} \vert \bar{f} \rangle=\langle i \vert i \rangle=1  $. Using such the completeness one can write down a unit operator at $ t=0 $ in the form $\hat{\textbf{1}}_{f,\hat{f}}= \vert f \rangle \langle f \vert + \vert \bar{f} \rangle \langle \bar{f} \vert  $. This enables one to introduce a vector of state $ \vert \bar{i} \rangle=\vert f \rangle \langle f \vert \bar{i} \rangle + \vert \bar{f} \rangle \langle \bar{f}\vert \bar{i} \rangle $ being "complementary"  to $ \vert i \rangle $, though not orthogonal to the latter i.e. $ \langle \bar{i} \vert i \rangle =\langle i \vert \bar{i} \rangle \neq 0$ . One can always parametrize the latter decompositions as follows $\vert i \rangle=cos \theta \vert f \rangle + sin \theta\vert \bar{f} \rangle$ and $\vert \bar{i} \rangle= cos \theta \vert f \rangle - sin \theta\vert \bar{f} \rangle$, in the same fashion as it was done for the ensemble of $ N $ spin-1/2 electrons in AAV-paper of Ref.[1]. Evidently, for such the case one has $ \vert f \rangle=\vert\uparrow \rangle $ and $ \vert \bar{f} \rangle=\vert\downarrow \rangle $, while  $ \langle f \vert i \rangle =cos\theta $ and $ \langle \bar{f} \vert i \rangle =sin \theta $, where constant parameter $ \theta $ represents the angle of spin polarization in the $ xz $-plane for the initial (or pre-selected) quantum state of each electron in the incident beam \cite{1}. It is evident, that $ \langle \bar{i} \vert f \rangle=\langle f \vert \bar{i} \rangle = \langle \bar{f} \vert i \rangle$ and $ \langle \bar{f} \vert \bar{i} \rangle=\langle \bar{i} \vert \bar{f} \rangle = -\langle i \vert \bar{f} \rangle=-\langle \bar{f} \vert i \rangle$.

Now let us introduce following general unit operator (projector) $ \hat{\textbf{1}}_{f,\bar{f}}(t) $ which acts as a unit operator at arbitrary instant of time $ t $, e.g. $ \hat{\textbf{1}}_{f,\bar{f}}(t)\vert f(t) \rangle=\vert f(t) \rangle $ and $ \hat{\textbf{1}}_{f,\bar{f}}(t)\vert \bar{f}(t) \rangle=\vert \bar{f}(t) \rangle $ in a following way

\begin{eqnarray} \nonumber
\hat{\textbf{1}}_{f,\hat{f}}(t)=\tilde{U}_{s}^{+}(t) \left( \vert f \rangle \langle f \vert + \vert \bar{f} \rangle \langle \bar{f} \vert \right)\tilde{U}_{s}(t) \nonumber \\
=  \tilde{U}_{s}^{+}(t) \frac{\left( \vert i \rangle  + \vert \bar{i} \rangle  \right)}{2\langle f \vert i \rangle} \langle f \vert \tilde{U}_{s}(t) + \tilde{U}_{s}^{+}(t) \frac{\left( \vert i \rangle  - \vert \bar{i} \rangle \right)}{2\langle \bar{f} \vert i \rangle} \langle \bar{f} \vert \tilde{U}_{s}(t) \nonumber 
\end{eqnarray}
\begin{equation}
\label{8}
\end{equation}
With the help of Eq.(9) one can rewrite decomposition (7) as follows

\begin{eqnarray} \label{9}
\frac{\langle f \vert \tilde{U}_{A}(t)\vert i\rangle}{\langle f \vert i\rangle}= 1 -i\int_{0}^{t}dt_{1} \frac{\langle f(t_{1})\vert \tilde{H}_{A}(t_{1}) \vert i \rangle}{\langle f \vert i \rangle} \nonumber \\
+ (-i)^{2} \int_{0}^{t}dt_{1} \int_{0}^{t_{1}}dt_{2} \frac{\langle f(t_{1}) \vert \tilde{H}_{A}(t_{1}) \hat{\textbf{1}}_{f,\bar{f}}(t_{2})\tilde{H}_{A}(t_{2}) \vert i(t_{2}) \rangle}{\langle f \vert i \rangle}  \nonumber \\
+ \ldots +(-i)^{n} \int_{0}^{t}dt_{1} \ldots \nonumber \\
 \int_{0}^{t_{n-1}}dt_{n} \frac{\langle f(t_{1})\vert \tilde{H}_{A}(t_{1})\hat{\textbf{1}}_{f,\bar{f}}(t_{2}) \ldots \hat{\textbf{1}}_{f,\bar{f}}(t_{n})\tilde{H}_{A}(t_{n}) \vert i(t_{n}) \rangle}{\langle f \vert i \rangle}\nonumber \\
+\ldots \nonumber \\
\end{eqnarray}  

At this point from Eqs.(5,10) it becomes evident, that in order to reduce propagator (8) to its \textit{weak quantum  evolution} form of Eq.(6) with following desired form of the propagator $ \langle f \vert \tilde{U}_{A}(t)\vert i\rangle / \langle f \vert i\rangle $

\begin{eqnarray} \label{10}
\frac{\langle f \vert \tilde{U}_{A}(t)\vert i\rangle}{\langle f \vert i\rangle} \cong \langle i \vert \tilde{U}_{A}(\hat{\Pi}_{f};t)\vert i\rangle_{w} \nonumber \\
= 1 -i\int_{0}^{t}dt_{1} \frac{\langle f(t_{1})\vert \tilde{H}_{A}(t_{1}) \vert i \rangle}{\langle f \vert i \rangle} + \nonumber \\
  (-i)^{2} \int_{0}^{t}dt_{1} \int_{0}^{t_{1}}dt_{2} \frac{\langle f(t_{1}) \vert \tilde{H}_{A}(t_{1})\vert i(t_{1}) \rangle \langle f(t_{2}) \vert \tilde{H}_{A}(t_{2}) \vert i \rangle}{(\langle f \vert i \rangle)^{2}}  \nonumber \\
+ \ldots +(-i)^{n} \int_{0}^{t}dt_{1} \ldots \nonumber \\
 \int_{0}^{t_{n-1}}dt_{n} \frac{\langle f(t_{1})\vert \tilde{H}_{A}(t_{1})\vert i(t_{1}) \rangle \ldots \langle f(t_{n}) \vert\tilde{H}_{A}(t_{n}) \vert i \rangle}{(\langle f \vert i \rangle)^{n}} \nonumber \\
+\ldots  \nonumber \\
\end{eqnarray}

one should be able to neglect $ \vert \bar{f} \rangle \langle \bar{f} \vert $ - component in the decomposition (9) of our projector $ \hat{\textbf{1}}_{f,\bar{f}}(t) $ being substituted into the decomposition (10); another requirement of the validity of Eq.(11) is the same (zero) phase of two time-dependent amplitudes $ \langle i \vert i(t) \rangle = \langle i \vert \tilde{U}_{s}(t) \vert i \rangle $ and $ \langle i \vert \bar{i}(t) \rangle=\langle \bar{i} \vert \tilde{U}_{s}(t) \vert \bar{i} \rangle $ meaning that one can also approximately neglect the $  \vert \bar{i} \rangle \langle f \vert$ contributions to the series (10). Obviously, the first among two latter requirements can be formulated as the restriction

\begin{equation} \label{11}
N \ll \left\vert \frac{\langle i \vert \bar{f} \rangle}{\langle i \vert f \rangle} \right\vert 
\end{equation}

whereas the second one can be formulated as follows

\begin{equation} \label{12}
\arg \lbrace \langle \bar{i} \vert \tilde{U}_{s}(t) \vert \bar{i} \rangle \rbrace \cong \arg \lbrace \langle i \vert \tilde{U}_{s}(t) \vert i \rangle \rbrace \simeq 0,
\end{equation}
 
One can obtain relations (12,13) expanding a product of $ N $ expansions (10) in the formula (8) under assumption that the largest correction to expression (11) in the formula 

\begin{equation} \label{13}
\langle f_{N}(t) \vert \tilde{U}^{\otimes N}_{A}(t)\vert i_{N}\rangle_{w} \cong \left( \langle f(t) \vert i\rangle \right)^{N} \left( \langle i \vert\hat{U}_{A}(\hat{\Pi}_{f};t)\vert i\rangle_{w}\right)^{N} 
\end{equation}

should be much smaller than r.h.s. of Eq.(14).

Now it becomes clear that conditions (12,13) are compatible with each other and both result in the propagator of ensemble weak evolution of the type (6). Recalling that, according to our presumed definition $ \tilde{H}_{A}(t')=\hat{H}_{A}(t')/N $ and $ \hat{H}_{A}(t') \propto \hat{A} $ one immediately obtains 

\begin{eqnarray}
\label{14}
\left(\langle i \vert\hat{U}_{A}(\hat{\Pi}_{f};t)\vert i\rangle_{w} \right)^{N} \cong \exp \left( -i \int_{0}^{t}dt' \frac{\langle i \vert \hat{\Pi}_{f} \hat{H}_{A,I}(t')\vert i \rangle}{\langle i \vert \hat{\Pi}_{f} \vert i \rangle}\right). \nonumber \\
\end{eqnarray} 

where $ \hat{H}_{A,I}(t)=\tilde{U}_{s}(t)\hat{H}_{A}(t)\tilde{U}_{s}^{+}(t)  $ - is the Hamiltonian of weak measurement in the interaction representation with respect to Hamiltonian $ \tilde{H}_{s}(t) $ in the evolution operator $ \tilde{U}_{s}(t) $ associated with certain strong measurement (or post-selection) on the ensemble of given identical quantum systems. Notice, that in order to "compactificate" series (11) into the exponential function of r.h.s. of Eq.(15) one needs one additional assumption $ \langle i(t_{1}) \vert i(t_{2}) \rangle \cong \langle i(t_{1}) \vert i(t_{1}) \rangle $ which is, however, just a consequence of the above derived restriction of Eq.(13). 

Therefore, Eq.(14) can be rewritten in the form 

\begin{equation} \label{15}
\langle f_{N}(t) \vert \tilde{U}^{\otimes N}_{A}(t)\vert i_{N}\rangle_{w} \cong \left( \langle f(t) \vert i\rangle \right)^{N} \exp \lbrace-i \varepsilon_{w}(t) q A_{w} \rbrace
\end{equation}

with standard weak value $ A_{w}=\frac{\langle f \vert \hat{A} \vert i \rangle}{\langle f \vert i \rangle} $ of operator $ \hat{A} $ (see Ref.[1]). 

Here it is interesting to note that inequality being opposite to the limit (12), namely, $ N \ll \left\vert \langle i \vert f \rangle / \langle i \vert \bar{f} \rangle \right\vert  $ - should represent a condition of the validity of so-called "null-weak values" approach developed in Ref.\cite{10} as an alternative which complements description via  weak values. At the same time, the above derivation clearly shows that in the absence of such strong inequalities as one of Eq.(12), i.e. in all cases where $ N \gtrsim \left\vert \langle i \vert \bar{f} \rangle / \langle i \vert f \rangle \right\vert  $ - for weak values, i.e. for very large statistical ensembles of identical quantum systems (and/or for not very small overlaps between pre- and post-selected quantum states), overall weak quantum evolution (and, hence, the weak values-) approach to temporal evolution of each quantum system in such ensemble - breaks down (see also related comments in the next section below).   

From Eqs.(9-16) one can notice that the entire description of quantum dynamics caused by weak measurement with post-selection in terms of weak values or, equally, in terms of weak quantum evolution of the system  can be valid only under special restrictions (12,13). For instance, in the case of parametrization being used in AAV-model of Ref.[1] the conditions (12,13) connect the size $ N $ of the ensemble of 1/2-spin electrons, the angle $ \theta $ of pre-selected 1/2- spin polarization in the $ xz $ -plane for each particle of the ensemble together with the "strength" $ \varepsilon_{st}(t)q_{x} $ of strong measurement associated with post-selection process in the double Stern-Gerlach type of experiment considered in Ref.[1]. Namely, conditions (12,13) in terms of this parametrization result in remarkable inequality 

\begin{equation} \label{16}
\varepsilon_{st}(t)q_{x} \sin (2\theta) \ll N \ll \tan(\theta).
\end{equation}
which should restrict the limits of the applicability  of overall weak values approach from Ref.\cite{1}. Also taking into account the spread renormalization $ \Delta_{x,z;(N)} = \Delta_{x,z}\sqrt{N} $ for the initial distribution of the ensemble of $ N $ spin-1/2 particles (each with distribution of the spread $ \Delta_{x,z} $) in the $ (x,z) $ plane in the double Stern-Gerlach experiment from Ref.\cite{1} one can derive from Eq.(17) following remarkable condition 

\begin{equation} \label{17}
\Delta_{x,z}\sqrt{\varepsilon_{st}(t)q_{x} \sin (2\theta)} \ll \Delta_{x,z;(w)}  \ll \Delta_{x,z}\sqrt{\tan(\theta)}.
\end{equation}

on the spread $ \Delta_{x,z;(w)}= \Delta_{x,z;(N)} $ of initially prepared ensemble which considerably restricts the existence of weak values $ S_{z}^{(w)} $ of the $ z $ component of 1/2-spin in the measurement outcome of double Stern-Gerlach experiment from AAV paper of Ref.\cite{1}. Therefore, if in such the experiment one has initially prepared beam of $ N $ identical particles with the spread $ \Delta_{x,z;(N)}  $ beyond the limits of inequality (18) then corresponded quantum evolution of such the ensemble in the process of two consequent (weak and strong) measurements \textit{will not be one of a weak type and hence it will not result in weak values in the measurement outcome}.
Evidently, constraints (12,13) and (17,18) signal about the limits of the applicability of entire weak values- and weak quantum evolution approach. One can see that latter approach remains valid only for high enough asymmetry in the probability amplitudes of pre- and post-selected quantum states (as it takes place in AAV-model \cite{1}) and only for finite-sized statistical ensembles of identical quantum systems under consideration (see also the related analysis in below).

\section{Measurement-determined quantum transition probability}

 Further, with respect to identities $ \vert f \rangle \langle f \vert = \hat{\Pi}_{f} $ and $ \langle i \vert f \rangle \langle f \vert i \rangle = \langle i \vert  \hat{\Pi}_{f} \vert i \rangle  $ one can rewrite Eq.(11) in a following way

\begin{eqnarray} \label{17}
 \langle i \vert \tilde{U}_{A}(\hat{\Pi}_{f};t)\vert i\rangle_{w}=\langle i \vert \tilde{U}_{A}(\hat{\Pi}^{w}_{f};t)\vert i\rangle \nonumber \\
 = 1 -i\int_{0}^{t}dt_{1} \langle i \vert \tilde{\Pi}^{w}_{f} \tilde{H}_{A,I}(t_{1}) \vert i \rangle + \nonumber \\
  (-i)^{2} \int_{0}^{t}dt_{1} \int_{0}^{t_{1}}dt_{2} \langle i \vert \tilde{\Pi}^{w}_{f} \tilde{H}_{A,I}(t_{1})\vert i \rangle \langle i \vert \tilde{\Pi}^{w}_{f} \tilde{H}_{A,I}(t_{2}) \vert i \rangle \nonumber \\
+ \ldots +(-i)^{n} \int_{0}^{t}dt_{1} \ldots \nonumber \\
 \int_{0}^{t_{n-1}}dt_{n} \langle i \vert \tilde{\Pi}^{w}_{f} \tilde{H}_{A,I}(t_{1})\vert i \rangle \ldots \langle i \vert \tilde{\Pi}^{w}_{f} \tilde{H}_{A,I}(t_{n}) \vert i \rangle \nonumber \\
+\ldots  \nonumber \\
\end{eqnarray}
with $ \tilde{H}_{A,I}(t)=\tilde{U}_{s}(t)\tilde{H}_{A}(t)\tilde{U}_{s}^{+}(t) $. 

In Eq.(19) I have introduced a new quantity, the projector 

\begin{equation} \label{18}
\tilde{\Pi}^{w}_{f}=\frac{\hat{\Pi}_{f}}{\langle i \vert \hat{\Pi}_{f} \vert i \rangle}
\end{equation}

which, evidently, can be treated as the \textit{operator of weak conditioned measurement}, or, alternatively, as the operator of \textit{strong fluctuative measurement}. These two definitions might seem controversial but one can easily justify the validity of both. 

Indeed, on one hand, operator (20) depends on both pre-and post-selected quantum states (i.e. on $ \vert i \rangle $ and $ \vert f \rangle $, correspondingly) this means it describes a projection on the post-selected quantum state being conditioned to  chosen pre-selected quantum state of the system. On the other hand, evident properties $ \langle i\vert \tilde{\Pi}_{f} \vert i \rangle = \langle i \vert i \rangle =1 $ and $ \langle f \vert \tilde{\Pi}_{f} \vert f \rangle = \vert \langle f \vert f \rangle \vert^{2} / \vert \langle i \vert f \rangle  \vert^{2}=1 / P_{i \rightarrow f} $ mean that the result of operator $ \tilde{\Pi}^{w}_{f} $ action on the corresponded pre-selected state $ \vert i \rangle $ averaged over this pre-selected state is the same as the result of scalar product $ \langle i \vert i \rangle =1 $ without any post-selection, i.e. the measurement described by the operator $ \tilde{\Pi}^{w}_{f} $ remains \textit{non-invasive} for the state $ \vert i \rangle $ "in average". In this sense related projective measurement (20) is "\textit{weak}". However, the operator $ \tilde{\Pi}^{w}_{f} $ by its definition, transforms vectors of state $ \vert i \rangle $ and  $ \vert f \rangle $ to vectors $ \vert f \rangle / \langle i \vert f \rangle $ and $ \vert f \rangle / \vert \langle i \vert f \rangle  \vert^{2} $, correspondingly, which both tend to $ \vert i \rangle $ if $ \langle i \vert f \rangle \rightarrow 1 $  and both diverge when $ \langle i \vert f \rangle \rightarrow 0 $ signalling about the appearance of strong quantum fluctuations when one tries to project certain initial (pre-selected) quantum state  $ \vert i \rangle $ to another (post-selected) quantum state  $ \vert f \rangle $ being approximately orthogonal to the former. Naturally, the latter situation happens when two near-orthogonal vectors $ \vert i \rangle $ and $ \vert f \rangle $ correspond to two non-commuting variables and their evolution is governed by two non-commuting Hamiltonians (corresponding to $ \tilde{H}_{A}(t) $ and $ \tilde{H}_{s}(t) $ ). Therefore, one can think of the operator (20) as of the \textit{operator of simultaneous measurement of two non-commuting variables}. In particular, a variable associated with quantum state $ \vert i \rangle $  remains well-defined in average after the projection defined by Eq.(20) onto another non-orthogonal quantum state  $ \vert f \rangle $, while a simultaneous measurement of the latter (i.e. of $ \vert f \rangle $) would give a strongly fluctuating result $ \langle f \vert \tilde{\Pi}_{f} \vert f \rangle = 1 / \vert \langle i \vert f \rangle  \vert^{2}\rightarrow \infty $ when $ \langle i \vert f \rangle  \rightarrow 0 $ if $ [\tilde{H}_{A}(t), \tilde{H}_{s}(t)] \neq 0 $. This fact, in turn, explains why according to Eqs.(12,13,16,17) for large enough ensembles, in the limit $ N \rightarrow \infty$, the picture of temporal evolution (19) described by weak values breaks down. The fact is that the limit $ N \rightarrow \infty$ describes the situation  $ [\tilde{H}_{A}(t),\tilde{H}_{s}(t)]\rightarrow 0 $, where variables corresponded to vectors $ \vert i \rangle $ and $ \vert f \rangle $ become effectively "classical" and, hence, strong quantum fluctuations during simultaneous measurement of both quantum states - disappear since these fluctuations average to zero due to extra contributions to average (5) which cannot be captured by the approach of "weak evolution" of Eqs.(15,19).

Now one can see a deep quantum statistical sense of formula (4) for transition probability between two arbitrary quantum states. Taking into account that from Eqs.(4,11,19)  $  \langle i \vert \tilde{U}_{A}(\hat{\Pi}_{f};t)\vert i\rangle_{w}= \langle i \vert \tilde{U}_{A}(\hat{\Pi}^{w}_{f};t)\vert i\rangle$- describes quantum evolution of the state $ \vert i \rangle $ with Hamiltonian $ \tilde{H}_{A}(t) $ accompanied by \textit{all possible combinations of strong fluctuative measurements of the post-selected quantum state $ \vert f \rangle $} one yields

\begin{equation} \label{19}
P_{i\rightarrow f}(t) =\vert \langle f(t) \vert i \rangle  \vert^{2}= \frac{\langle i \vert \hat{\Pi}_{f} \hat{U}_{A}(t) \vert i\rangle}{\langle i \vert\hat{U}_{A}(\hat{\Pi}^{w}_{f}, t)\vert i\rangle}
\end{equation} 

for any two chosen pre- and post-selected quantum states ($  \vert i \rangle $ and $  \vert f(t) \rangle $), arbitrary evolution operator $ \hat{U}_{A}(t) $ of weak measurement between these states and weak quantum evolution determined by Eq.(19) interrupted in all possible moments of time by strong fluctuative- (or weak-conditioned) measurements, one could perform by means of operator (20). For  the  ensembles of identical quantum systems the limits of the applicability of formula(21) are defined by means of restrictions (12,13) on the size of the related statistical ensemble, initial and final quantum states overlap and the strength of post-selecting "strong" measurement.

One can conclude from all the above, that equation (21) means that \textit{any transition probability from arbitrary quantum state $ \vert i \rangle $ to another arbitrary quantum state $ \vert f(t) \rangle $ can be calculated, in principle, as the ratio of two time-propagators corresponded to two different types of system's evolution between pre- and  post-selected quantum states of interest.} Namely, the numerator of Eq.(21) represents time-propagator describing quantum evolution of the state $ \vert i \rangle $ due to certain weak measurement governed by arbitrary Hamiltonian $ \tilde{H}_{A,I}(t) $ with consequent post-selection of the chosen quantum state $ \vert f \rangle $ via arbitrary strong measurement. Whereas, the denominator of Eq.(21) describes \textit{weak quantum evolution} of the system where a conventional time-evolution with $ \tilde{H}_{A,I}(t) $ gets interrupted by \textit{all possible combinations} of strongly fluctuating projective measurements (20) of the post-selected quantum state $ \vert f(t) \rangle $ performed at \textit{all possible stages} of system's temporal evolution. 

Now, according to the common statistical meaning of quantum transition probability between two arbitrary (initial and final) quantum states $ \vert i \rangle $  and $ \vert f(t) \rangle $ , the ratio (21) should be thought of as \textit{the number of "successful" system "histories" (or trajectories in the time domain) all resulting in just one successful measurement of post-selected quantum state $ \vert f(t) \rangle $ on all members of the ensemble at final point $ t $ of ensemble's "history" - divided by the total number of system's "histories" (or equally, time-trajectories) those ended at $ t $ and contained all possible strong fluctuative- (or equally weak conditioned) successful measurements of this post-selected quantum state $ \vert f \rangle $ on all ensemble's members at all possible stages of system's weak quantum evolution}. Such a representation via weak quantum evolution for arbitrary quantum transition probability becomes possible due to fine-tuned quantum fluctuations generated by all possible sequences of weak- and strong fluctuative measurements of two non-commutative observables each participating in given quantum transition of the system from its initial to final quantum state. 

As the consequence of the above context, one can claim that \textit{weak quantum evolution concept} (with  weak values of corresponded observable) being introduced for arbitrary quantum system (or finite ensemble of identical quantum systems) \textit{describes a fixed class of quantum evolutions (or quantum trajectories) of a given quantum system, where certain (post-selected) quantum state appears most frequently as a result of related strong fluctuative- (or weak conditioned) measurement during the time interval of overall quantum  evolution of the system}. This consequence is seemed to have some parallels with semi-classical trajectories approach to weak values developed earlier in Ref. \cite{12}. Also here it is worth to point out on the interesting connection has been revealed in Ref.\cite{13} between quantum Berry phase of target quantum system during its evolution from pre- to post-selected quantum state and a measurable weak value. In the above framework such a connection could serve as one among fingerprints of system's weak quantum evolution.

The latter statements, in turn, mean that \textit{weak quantum evolution and related weak values of quantum observables are not just abstract things or artefacts of quantum calculation, but rather these terms reflect real strongly correlated quantum effects in system's temporal evolution accompanying weak measurements on finite statistical ensembles of identical quantum systems with identical pre- and post-selection. The above theory shows that such the effects should accompany any quantum evolution of any quantum system from its arbitrary initial to arbitrary (non-orthogonal) final quantum state, in order to match a conventional probabilistic meaning of quantum transition probability, which is prescribed \textit{a-priori} to a given quantum system with chosen initial and final quantum states}.

\section{Conclusions}

In the above it was shown that quantum weak values are, in fact, fingerprints of a more general novel quantum phenomenon of weak quantum evolution for arbitrary quantum system with pre- and post-selection being valid in the presence of weak measurement in the process of system's evolution in the case of small overlap between its pre- and post-selected quantum states, when such quantum system is either unique or is a part of finite (and not very large) statistical ensemble of identical quantum systems (with identical pre- and post-selections). In the above the concept of weak quantum evolution has been introduced for the first time and explained qualitatively. The limits of the applicability for weak quantum evolution regime - and, hence, those for weak values have been established. In particular, it has been found that revealed weak regime of quantum evolution resulting in weak values can take place only for finite (being not very large) statistical ensembles of identical quantum systems with fixed inequality between the size of ensemble and overlap of pre-and post-selected quantum states. It was shown, that such conditions of weak quantum evolution regime should be applied to any realistic examples of a model quantum ensemble including one from the seminal Aharonov-Albert-Vaidman paper  where the concept of quantum operator weak values had been introduced for the first time. As the result, it has been demonstrated that the probability of quantum transition between two arbitrary quantum states of the system can be defined via introduced weak quantum evolution of a given quantum system between those two quantum states. The latter fact means that weak quantum evolution is real for any quantum system: it governs quantum fluctuations in any quantum system in order to fulfil the statistical meaning of the probability of quantum transition from chosen initial to final state for given quantum system. The above study  is practically important because of the fast growing experimental background in the area of weak measurements and weak values. On the other hand, weak quantum evolution concept intimately connects the fact of weak values existence with the existence of a conventional quantum transition probability from pre-selected to post-selected quantum state. The latter connection represents a brand new angle of view on the entire weak values problem, which is able to shed more light on further interconnections in the body of conventional quantum mechanics.

\section{Acknowledgement} Author thanks to Lev Vaidman for providing the reference to his recent experimentally verified research on weak values. Author thanks also to Yutaka Shikano for providing the  references to his paper on the connection between Berry phase and weak values. This study was partially supported by the research grant No.09/01-2021(2) from the National Academy of Sciences of Ukraine. 

\end{normalsize}


\begin{references}
\bibitem{9} In the AAV-paper for $ \hat{H}_{s}(t) $ one has $ q_{x}=x $,  $ d\varepsilon_{st}(t)/dt= -\mu\left(\partial B_{x}/\partial x \right)g[y(t,\tau_{s})] $ with $ \partial B_{x}/\partial x $ - a gradient of the external magnetic field in $ x$-direction and  $ g[y(t,\tau_{s})] $ is the same type of "pulse" function of time as $ g[y(t,\tau_{w}] $ in the main text, but the former is dependent on $ \tau_{s} $ instead of $ \tau_{w} $, where $ \tau_{s}=\tau - \tau_{w} $ - is electron's "time of flight" in $ y $-direction in the region of inhomogenious $x $-component of magnetic field, while the time $ \tau $($ \tau >\tau_{s,w} $) - represents total "time of flight" of the electron along $ y $-direction from the source to one of the (post-selected) screens.
\end{references}
\end{document}